\documentstyle[amssymb,aps]{revtex}

\begin{document}
\author{R.A. Serota\thanks{%
serota@physics.uc.edu} and B. Goodman\thanks{%
goodman@physic.uc.edu}}
\address{Department of Physics\\
University of Cincinnati\\
Cincinnati, OH\ 45221-0011}
\title{Quantum absorption in small metal particles}
\date{07/28/98}
\maketitle

\begin{abstract}
We evaluate the electric dipole absorption in small metal particles in a
longitudinal electric field taking into account the Fermi-Thomas screening.
When either the level broadening or the frequency of the field are larger
than the mean energy-level spacing, the main contribution to absorption is
classical, with quantum corrections. When both the broadening and the
frequency are smaller than the mean level spacing, the absorption is
manifestly quantum and can be understood in terms of the two-level system.
\end{abstract}

\section{Introduction}

The classical absorption in small diffusive metal particles and narrow films
has been evaluated in a companion paper\cite{SG}, which hereafter is
referred to as I. It was found that, aside from the corrections of the order 
$\left( \Lambda L\right) ^{-1}$, where $\Lambda $ is the Thomas-Fermi
wave-vector, 
\begin{equation}
\Lambda ^{2}=4\pi e^{2}\frac{dn}{d\mu }  \label{Lambda}
\end{equation}
$dn/d\mu $ is the thermodynamic density of states and $L$ is the system
size, the result coincides with that\cite{LL} obtained using the (complex)
Drude dielectric constant, 
\begin{equation}
\epsilon _{D}=1-\frac{4\pi i\sigma _{D}}{\omega }  \label{epsilon_Drude}
\end{equation}
for metal characterization. Here 
\begin{equation}
\sigma _{D}=\frac{\sigma _{0}}{1+i\omega \tau }  \label{sigma_Drude}
\end{equation}
is the Drude conductivity, $\sigma _{0}=D\Lambda ^{2}/4\pi $ is the
Boltzmann conductivity, and $D$ and $\tau $ are, respectively, the electron
diffusion coefficient and scattering time.

Namely, for a spherical particle of radius $a$ such that $a\Lambda \gg 1$,
the absorption in the oscillating electric field of amplitude $E_{0}$ and
frequency $\omega \ll \tau ^{-1}$ is given by\cite{SG} 
\begin{equation}
Q_{class}=\frac{9\left( \omega E_{0}\right) ^{2}V}{2\left( 4\pi \right)
^{2}\sigma _{0}}\left( 1-\frac{11}{2a\Lambda }\right) =Q_{RD}\left( 1-\frac{
11}{2a\Lambda }\right)  \label{Q_class}
\end{equation}
where $V=4\pi a^{3}/3$ is the particle volume. In eq. (\ref{Q_class}), $%
Q_{RD}$ is the standard Rayleigh-Drude result for absorption where it is
assumed that the applied field is screened due to the surface charge. The
second term in parentheses describes the correction due to the fact that the
Thomas-Fermi screening occurs, in reality, within the layer $\sim \Lambda
^{-1}$ from the particle surface. In what follows, we will neglect such
corrections.

The subject of this work is the quantum limit of the electric-dipole
absorption. If the level broadening $\gamma $ or $\omega $ are much larger
that the mean energy-level spacing $\Delta $ then, barring small quantum
corrections, the absorption is classical and is given by eq. (\ref{Q_class}
). If, however, $\gamma ,\omega \lesssim $ $\Delta $, the absorption can be
described in terms of the two-level system and is manifestly quantum. We
evaluate the temperature dependence of quantum absorption using the ideas -
developed in relation to the original work by Gor'kov and Eliashberg\cite{GE}
- of Lushnikov, Simonov and Maksimenko\cite{LMS} and Shklovskii\cite{S}.
Below, we will assume that $\gamma <\omega \ll D/a^{2}$ - the inverse time
of the electron diffusion to the boundary, and neglect the difference
between $Vdn/d\mu $ and $\upsilon =2\Delta ^{-1}$ - the mean level density
at the Fermi level\footnote{$2$ accounts for spin degeneracy}.

\section{Classical absorption}

We refer to I for a detailed analysis of classical absorption. In summary,
eq. (\ref{Q_class}) was obtained as a result of solving the Maxwell equation
and the current continuity condition combined with the generalized Einstein
transport (constitutive) equation and using the appropriate boundary
condition. However, for the purposes which will be explained below, we
concentrate on the alternative, yet equivalent, linear-response formulation 
\cite{SG}, \cite{WM} wherein eq. (\ref{Q_class}) can be also obtained from
(see Appendix B in Ref.\cite{SG}) 
\begin{eqnarray}
Q_{class} &=&\frac{\omega ^{2}\Lambda ^{4}D}{2\left( 4\pi \right) ^{2}\sigma
_{0}}\int \int \phi _{st}\left( {\bf r}\right) d\left( {\bf r},{\bf r}%
^{\prime };0\right) \phi _{st}\left( {\bf r}^{\prime }\right) d{\bf r}%
^{\prime }d{\bf r}  \label{Q_class-2.1} \\
&=&\frac{\omega ^{2}E_{0}^{2}\Lambda ^{4}D}{2\left( 4\pi \right) ^{2}\sigma
_{0}}\frac{1}{3}\int \int f\left( r\right) \left( {\bf r\cdot r^{\prime }}%
\right) d{\bf \left( r,r^{\prime };0\right) }f\left( r^{\prime }\right) d%
{\bf r}^{\prime }d{\bf r}  \label{Q_class-2.2}
\end{eqnarray}
where $d\left( {\bf r},{\bf r}^{\prime };0\right) $ is the static limit of
the diffusion propagator (diffuson) satisfying the equation 
\begin{eqnarray}
D{\bf \nabla }^{2}d\left( {\bf r},{\bf r}^{\prime };\omega \right)
&=&-\delta ({\bf r}-{\bf r}^{\prime })+\frac{1}{V}+i\omega d\left( {\bf r},%
{\bf r}^{\prime };\omega \right)  \label{d_equ} \\
{\bf \nabla }_{n}d\left( {\bf r},{\bf r}^{\prime };\omega \right)
|_{\partial } &=&{\bf \nabla }_{n}^{\prime }d\left( {\bf r},{\bf r}^{\prime
};\omega \right) |_{\partial ^{\prime }}=0  \label{d_bc}
\end{eqnarray}
and $\phi _{st}$ is the (quasi) static potential inside the particle, 
\begin{equation}
\phi _{st}\left( {\bf r}\right) =\left( -{\bf E}_{0}\cdot {\bf r}\right)
f\left( r\right)  \label{phi_stat}
\end{equation}
The latter falls off exponentially within the layer $\sim \Lambda ^{-1}$
from the particle surface according to 
\begin{eqnarray}
f\left( r\right) &=&\frac{3a}{r}%
\mathop{\rm csch}%
\left( a\Lambda \right) i_{1}\left( r\Lambda \right)  \label{f} \\
i_{1}\left( x\right) &=&\sqrt{\frac{\pi }{2x}}I_{\frac{3}{2}}\left( x\right)
\label{i1}
\end{eqnarray}
where $i_{1}\left( x\right) $ and $I_{\frac{3}{2}}\left( x\right) $ are the
spherical Bessel and Bessel function of imaginary argument respectively. As
was already mentioned above, we neglect the corrections in orders of $\left(
a\Lambda \right) ^{-1}$ (the complete expressions can be found in I).

Solving eqs. (\ref{d_equ}) and (\ref{d_bc}), we find 
\begin{equation}
d\left( {\bf r},{\bf r}^{\prime };0\right) =\frac{1}{4\pi \left| {\bf r-r}%
^{\prime }\right| }+\frac{2r}{3V}+\sum_{l=1}^{\infty }\frac{\left(
l+1\right) \left( rr^{\prime }\right) ^{l}}{4\pi la^{2l+1}}P_{l}\left( \cos
\theta \right)  \label{d_solution}
\end{equation}
where $P_{l}\left( \cos \theta \right) $ is the Lagrange polynomial and $%
\theta $ is the angle between ${\bf r}$ and ${\bf r}^{\prime }$. Due to the
term ${\bf r\cdot r^{\prime }=}rr^{\prime }\cos \theta $, only the term $%
\propto P_{1}\left( \cos \theta \right) =\cos \theta $ in eq. (\ref
{d_solution}), 
\begin{equation}
d_{1}\left( {\bf r},{\bf r}^{\prime };0\right) =\left( \frac{1}{4\pi }\frac{%
r_{<}}{r_{>}^{2}}+\frac{rr^{\prime }}{2\pi a^{3}}\right) \cos \theta
\label{d1_solution}
\end{equation}
contributes to the integral of eq. (\ref{Q_class-2.2}), where $r_{<}$ and $%
r_{>}$ denote the lesser and greater, respectively, of $r$ and $r^{\prime }$%
. Evaluating the integral, we find the standard Rayleigh-Drude result for
absorption, $Q_{RD}$ in eq. (\ref{Q_class}).

We emphasize that all above results are obtained, strictly speaking, for $%
\ell \Lambda \ll 1$ (see I). The opposite limit $\ell \Lambda \gg 1$ must be
a subject of a separate analysis. In such a limit an analogy with the
anomalous skin-effect in a transverse field can be drawn wherein the
diffusive description of electrons is possible at a distance of order or
larger than $\ell $ from the boundary while the field penetration (skin)
depth is smaller than $\ell $\cite{Z},\cite{LL2}. In the longitudinal field
studied here, it is the Thomas-Fermi screening length that plays the role of
the field penetration depth. While the depth in the anomalous skin effect is
purely dynamical, the Thomas-Fermi depth is dominated by the static
component $\sim \Lambda ^{-1}.$

\section{Quantum absorption}

Quantum absorption can be evaluated using the standard time-dependent
perturbation theory\cite{LL3} where one must distinguish between the
transitions in the continuous and discrete spectra. The spectrum can be
regarded as effectively continuous when $\gamma \gg \Delta $ in which
circumstance (as well as for $\omega \gg \Delta $) it is classical and is
given by $Q_{RD}$, up to quantum corrections. When, on the other hand, $%
\gamma ,\omega \lesssim $ $\Delta $, the absorption is manifestly quantum
and is determined by the two-level physics.

\subsection{Continuous spectrum}

For a continuous spectrum, the transition probability per unit time from the
state $\left| i\right\rangle $ to the states $\left| f\right\rangle $ in the
interval of energies $d\varepsilon _{f}$ is given by ($\hbar =1$) 
\begin{equation}
dw_{if}=2\pi \left| F_{if}\right| ^{2}\delta \left( \varepsilon
_{f}-\varepsilon _{i}-\omega \right) \upsilon \left( \varepsilon _{f}\right)
d\varepsilon _{f}  \label{diff_w}
\end{equation}
where $\upsilon $ is the level density and $\upsilon \left( \varepsilon
_{f}\right) d\varepsilon _{f}$ is the number of final states in the energy
interval $d\varepsilon _{f}$. The matrix element $F_{if}$ corresponds to the
periodic perturbation of frequency $\omega $, 
\begin{equation}
\widehat{V}=\widehat{F}e^{-i\omega t}+\widehat{F}^{+}e^{i\omega t}
\label{V_hat}
\end{equation}
In the present case, $F=\overrightarrow{{\cal P}}{\bf \cdot E}_{0},$ where $%
\overrightarrow{{\cal P}}$ is the dipole moment of a single particle. Using
eq. (\ref{diff_w}) the absorption is found as, 
\begin{equation}
Q_{cont}=2\pi \int \int \left| F_{if}\right| ^{2}\left( \varepsilon
_{f}-\varepsilon _{i}\right) \delta \left( \varepsilon _{f}-\varepsilon
_{i}-\omega \right) \upsilon \left( \varepsilon _{i}\right) \upsilon \left(
\varepsilon _{f}\right) \left[ n\left( \varepsilon _{i}\right) -n\left(
\varepsilon _{f}\right) \right] d\varepsilon _{i}d\varepsilon _{f}
\label{Q_cont}
\end{equation}
where $\upsilon \left( \varepsilon _{i}\right) d\varepsilon _{i}$ is the
number of initial states in the interval $d\varepsilon _{i}$ and $n$ is the
Gibbs thermal occupancy of the state.

For electrons in the Fermi see, the integration reduces to the one over the
single-electron states $\left| i\right\rangle $ and $\left| f\right\rangle $
with the matrix element being that of a single-electron dipole moment $e%
\widehat{{\bf R}}$. Since each particle has a different impurity
configuration, we average over the disorder (denoted by the angular
brackets) to find the following expression for the mean value of absorption: 
\begin{equation}
\left\langle Q_{cont}\right\rangle =\frac{2\pi }{3}e^{2}E_{0}^{2}\int \int
\left| \widehat{{\bf R}}_{if}\right| ^{2}\left( \varepsilon _{f}-\varepsilon
_{i}\right) \delta \left( \varepsilon _{f}-\varepsilon _{i}-\omega \right)
\left\langle \upsilon \left( \varepsilon _{i}\right) \upsilon \left(
\varepsilon _{f}\right) \right\rangle \left[ f\left( \varepsilon _{i}\right)
-f\left( \varepsilon _{f}\right) \right] d\varepsilon _{i}d\varepsilon _{f}
\label{Q_cont_mean}
\end{equation}
where $f\left( \varepsilon \right) $ is the Fermi thermal occupancy factor.
Integration in eq. (\ref{Q_cont_mean}) is performed by changing the
integration variables to $y=\varepsilon _{f}+\varepsilon _{i}$ and $%
x=\varepsilon _{f}-\varepsilon _{i}$ yielding 
\begin{equation}
\left\langle Q_{cont}\right\rangle =\frac{2\pi }{3}\omega ^{2}e^{2}\upsilon
^{2}E_{0}^{2}\left| \widehat{{\bf R}}_{if}\right| ^{2}\frac{\left\langle
\upsilon \left( 0\right) \upsilon \left( \omega \right) \right\rangle }{%
\upsilon ^{2}}  \label{Q_cont_mean-2}
\end{equation}
where the mean position of the Fermi level is chosen to be at zero.

Notice that, in the usual manner\cite{GE},\cite{LMS}, we decoupled the
matrix element from the level-density correlation function. Moreover, the
former is evaluated in the semi-classical approximation, which is justified
for the large quantum numbers corresponding to the Fermi level\cite{GE},\cite
{LMS}. Finally, $\widehat{{\bf R}}$ must include the effect of screening of
the external field ${\bf E}_{0}$\cite{LMS}, 
\begin{equation}
\widehat{{\bf R}}=\frac{1}{2}f\left( r\right) \widehat{{\bf r}}
\label{R_operator}
\end{equation}
A more rigorous quantum-mechanical response-function formulation in Ref.\cite
{BM} confirms the validity of this procedure.

From the semi-classical expression\cite{GE},\cite{LMS} 
\begin{eqnarray}
\left| \widehat{{\bf R}}_{if}\right| ^{2} &=&\frac{1}{\pi \upsilon }\int
\int \left( {\bf R\cdot R^{\prime }}\right) d{\bf \left( r,r^{\prime
};0\right) }d{\bf r}^{\prime }d{\bf r}  \label{R_matrix_elem} \\
&=&\frac{1}{4\pi \upsilon }\int \int f\left( r\right) \left( {\bf r\cdot
r^{\prime }}\right) d{\bf \left( r,r^{\prime };0\right) }f\left( r^{\prime
}\right) d{\bf r}^{\prime }d{\bf r}  \label{R_matrix_elem-2}
\end{eqnarray}
and eqs. (\ref{Q_class-2.2}) and (\ref{Q_cont_mean-2}), it follows that 
\begin{equation}
\left\langle Q_{cont}\right\rangle =Q_{RD}\frac{\left\langle \upsilon \left(
0\right) \upsilon \left( \omega \right) \right\rangle }{\upsilon ^{2}}
\label{Q_cont-vs-Q_RD}
\end{equation}
showing that, aside from the quantum correction in the correlation function $%
\upsilon ^{-2}\left\langle \upsilon \left( 0\right) \upsilon \left( \omega
\right) \right\rangle $, the result obtained in this approximation coincides
with $Q_{RD}$. This is in agreement with Ref.\cite{LMS}.

The expression for $\left\langle \upsilon \left( 0\right) \upsilon \left(
\omega \right) \right\rangle $ for any value of level broadening $\gamma $
was derived in Ref.\cite{SZ}. For $\gamma \gg \Delta $, it is given by a
perturbation theory expression\cite{AS} 
\begin{equation}
\frac{\left\langle \upsilon \left( 0\right) \upsilon \left( \omega \right)
\right\rangle }{\upsilon ^{2}}=1+\frac{1}{\eta \pi ^{2}}%
\mathop{\rm Re}%
\frac{\Delta ^{2}}{\left( -i\omega +\gamma \right) ^{2}}  \label{nu_corr}
\end{equation}
where $\eta =1,2,4$ for Gaussian orthogonal (GOE), unitary (GUE) and
symplectic (GSE) ensembles, respectively. When $\gamma \lesssim $ $\Delta $,
the second term in the parentheses acquires an ensemble-specific oscillatory
behavior\cite{B-W/M/E}. However, for $\omega \gg \Delta $ it is still small
in the order $\Delta ^{2}/$ $\omega ^{2}$ which supports the validity of the
present approximation, even though the spectrum becomes discrete in this
limit.

Notice that one of the conditions specified earlier, $\omega \ll D/a^{2}$,
has been used for the diffusive evaluation of the dipole matrix element and
also in the use of the ''zero-mode'' approximation for the correlation
function (\ref{nu_corr}) where the kinetic terms $\sim D/a^{2}$ are omitted
relative to $\omega $. The other condition, $\gamma <\omega $, is needed to
justify the Fermi golden rule approximation for the evaluation of the
absorption in eqs. (\ref{Q_cont}) and (\ref{Q_cont_mean}). Namely, the
assumption of a sharp peak in the transition probability at $\varepsilon
_{f}-\varepsilon _{i}=\omega $ would not be meaningful otherwise. Notice
that, more rigorously, the $\delta $-function should have been replaced by $%
\gamma /\left[ \pi \left( \gamma ^{2}+\left( \varepsilon _{f}-\varepsilon
_{i}-\omega \right) ^{2}\right) \right] $. However, this would not effect
the parametric dependence of the quantum correction.

\subsection{Discrete spectrum}

By definition, the level density assumes averaging over many levels. When $%
\gamma ,\omega \lesssim $ $\Delta $, on the other hand, the energy spectrum
must be treated as discrete and the transitions leading to absorption occur
predominantly to the nearest level\cite{S}. This is in complete analogy with
the thermodynamical quantities which are evaluated using a few-level
approximation\cite{D-S/SS}. Therefore, one needs to use another descriptor
of level statistics, namely the probability $P\left( x\right) dx$ of finding
the next nearest level at the distance $x$ of a given level\footnote{%
Notice that if one formally extends the expression for the level-density
correlation function to $\omega <$ $\Delta $, the following relationship
exists\cite{B-W/M/E} between the former and the probability density $P$: $%
\lim_{\varepsilon _{f}-\varepsilon _{i}\rightarrow 0}\left( \left\langle
\upsilon \left( \varepsilon _{i}\right) \upsilon \left( \varepsilon
_{f}\right) \right\rangle \upsilon ^{-2}\right) =\lim_{\varepsilon
_{f}-\varepsilon _{i}\rightarrow 0}\left( P\left( \varepsilon
_{f}-\varepsilon _{i}\right) \Delta \right) $}. Furthermore, as was pointed
out by Shklovskii\cite{S}, unless $T\gg $ $\Delta $, the thermal populations
are those of the two-level system rather than those given by the Fermi
distribution function.

Taking into account all these considerations, we shall consider the
following cases (ignoring $\gamma $ in what follows and assuming that eq. (%
\ref{R_matrix_elem-2}) is still valid):

\subsubsection{$T\gg $ $\Delta $}

In this case, 
\begin{equation}
\left\langle Q_{disc}^{\left( 1\right) }\right\rangle =\frac{2\pi }{3}%
e^{2}E_{0}^{2}\left| \widehat{{\bf R}}_{if}\right| ^{2}\upsilon \Delta
\sum_{i}\int x\delta \left( x-\omega \right) P\left( x\right) \left[ f\left(
\varepsilon _{i}\right) -f\left( \varepsilon _{i}+x\right) \right] dx
\label{Q_disc_1_sum}
\end{equation}
where summation is over all nearest-level pairs within $\sim T$ of the Fermi
level. Replacing it by integration, $\sum_{i}\rightarrow \upsilon \int
d\varepsilon _{i}$, denoting $x=\left( \varepsilon _{f}-\varepsilon
_{i}\right) $ and performing integration on $y=\varepsilon _{f}+\varepsilon
_{i}$ first, we obtain 
\begin{eqnarray}
\left\langle Q_{disc}^{\left( 1\right) }\right\rangle &=&\frac{2\pi }{3}%
e^{2}\upsilon ^{2}E_{0}^{2}\left| \widehat{{\bf R}}_{if}\right| ^{2}\Delta
\int \int x\delta \left( x-\omega \right) P\left( x\right) \left[ f\left(
\varepsilon _{i}\right) -f\left( \varepsilon _{f}\right) \right]
d\varepsilon _{i}dx  \label{Q_disc_1-1} \\
&=&\frac{2\pi }{3}e^{2}\upsilon ^{2}E_{0}^{2}\left| \widehat{{\bf R}}%
_{if}\right| ^{2}\Delta \int x^{2}\delta \left( x-\omega \right) P\left(
x\right) dx=Q_{RD}\Delta P\left( \omega \right)  \label{Q_disc_1-2}
\end{eqnarray}
For GOE, for instance, this implies $\left\langle Q_{disc}^{\left( 1\right)
}\right\rangle \sim \omega ^{3}$.

\subsubsection{$T<$ $\Delta $}

In this case, the only transitions of significance are those from the
(occupied) Fermi level to the next (unoccupied) level. The implication of
the latter is two-fold. First, there is no summation over the initial states%
\cite{S}, $\sum_{i}$. Second, the thermal factors are those of the two-level
system, 
\begin{equation}
f\left( 0\right) =\frac{1}{1+e^{-\frac{x}{T}}}\text{, }f\left( x\right) =%
\frac{e^{-\frac{x}{T}}}{1+e^{-\frac{x}{T}}}  \label{f_two_level}
\end{equation}
Consequently, we find the following expression for the absorption 
\begin{eqnarray}
\left\langle Q_{disc}^{\left( 2\right) }\right\rangle &=&\frac{2\pi }{3}%
e^{2}E_{0}^{2}\left| \widehat{{\bf R}}_{if}\right| ^{2}\upsilon \Delta \int
x\delta \left( x-\omega \right) P\left( x\right) \left[ f\left( 0\right)
-f\left( x\right) \right] dx  \label{Q_disc_2-1} \\
&=&Q_{RD}\upsilon ^{-1}\Delta \omega P\left( \omega \right) \tanh \left( 
\frac{\omega }{2T}\right)  \label{Q_disc_2-2} \\
&\simeq & 
\begin{array}{l}
Q_{RD}\frac{\Delta }{2\upsilon T}P\left( \omega \right) \text{,} \\ 
\\ 
Q_{RD}\frac{\Delta }{\upsilon }\omega P\left( \omega \right) \text{,}
\end{array}
\begin{array}{l}
\omega <T \\ 
\\ 
T<\omega
\end{array}
\label{Q_disc_2_limits}
\end{eqnarray}
For the orthogonal ensemble\cite{B-W/M/E}, the second of eq. (\ref
{Q_disc_2_limits}) implies that 
\begin{equation}
\left\langle Q_{disc}^{\left( 2\right) }\right\rangle =\frac{\pi ^{2}}{12}%
Q_{class}  \label{Q_disc_2_limit_2_vs_Q_class}
\end{equation}
The temperature dependence of the quantum absorption given by eqs. (\ref
{Q_disc_1-2}) and (\ref{Q_disc_2_limits}) has been described in Ref.\cite{S}.

\section{Conclusions}

We have argued that the quantum effects on electric-dipole absorption are
significant only when $\omega <$ $\Delta $. The main assumption of this
derivation is that the matrix element of the dipole moment can be evaluated
classically, in accordance with eq. (\ref{R_matrix_elem-2}), where the
Thomas-Fermi screening is taken into account as well. The quantum effects
are manifested through the use of the probability distribution function for
the energy spacing to the next-nearest level and the use of thermal
occupation factors of the two-level systems.

As was pointed out in I, since there is no screening for the magnetic-dipole
absorption, in classical electrodynamics the latter becomes dominant for
very small particle sizes. It is, therefore, important, to investigate the
quantum effects on the magnetic-dipole absorption. This will be done in a
future publication.

\section{Acknowledgments}

This work was not supported by any funding agency.

\end{document}